An in- and ex-situ TEM study into the oxidisation of titanium (IV) sulphide

Edmund Long[1,2*], Sean O'Brien[1,2], Edward A. Lewis[3], Eric Prestat[3], Clive Downing[2], Clotilde S. Cucinotta[1], Stefano Sanvito[1,2], Sarah J. Haigh[3], and Valeria Nicolosi[1,2,3,4]

[1] School of Physics, Trinity College Dublin, Dublin 2, Ireland
[2] CRANN and AMBER Research Centres, Dublin, Dublin 2, Ireland
[3] School of Materials, The University of Manchester, Oxford Road, Manchester, M13 9PL, United Kingdom
[4] School of Chemistry, Trinity College Dublin, Dublin 2, Ireland
*Email: longed@tcd.ie

Introduction

Layered materials such as graphite and some clays had been used in bulk form for hundreds of years prior to the isolation of graphene in 2004[1], within fields as diverse as agriculture[2], art[3], and construction[4] whilst also commonly used as dry lubricants. However, it was this discovery by Geim and Novoselov which opened the doors for research into these materials on the nanoscale, and the novel properties which became accessible. A layered material has strongly covalently-bonded planes of material held together by weaker van der Waals interactions. These planes can be separated through exfoliation methods ranging from mechanical exfoliation (the 'scotch-tape' method) for producing a few flakes at a time[1], to liquid phase exfoliation capable of producing hundreds of litres of dispersion-containing flakes within an hour[5]. The dispersions produced by liquid exfoliation methods can then be used in a range of scalable industrial processes.

Graphene was merely the starting point, however, soon other layered materials started to be exfoliated down to monolayers, sometimes changing their properties drastically[6]. Transition metal dichacogenides (TMDs) are one such category, typically consisting of a layer of transition metal atoms sandwiched between two layers of chalcogenide. They first became of interest when it was discovered that the bandgap of molybdenum disulphide ($MoS_2$) underwent a shift from indirect in the bulk to direct as a monolayer, with the gap itself tuneable based on the number of layers[7].

The focus of this paper is on titanium (IV) sulphide, $TiS_2$. High energy and power density mean $TiS_2$ is the focus of much interest within the energy-storage community as a battery cathode material[8], whilst it has also been used as a catalyst upon which Pt and Au water-splitting catalysts were grown[9]. It has also been proposed that alloying with $TiO_2$ would reduce the bandgap to optimise photon absorption for photocatalytic water splitting[10]. The exact bandgap of $TiS_2$ is hard to specify, as a range of values have been put forward across the literature, ranging from -1.5 to 2.5 eV, but it is usually considered to be either a semi-metal (negative bandgap arising from overlapping conduction and valence bands) or a small-gap semiconductor[11–13]. One possible explanation for this range of values is that the samples of $TiS_2$ being measured had already undergone some small degree of oxidisation, or that samples differed stoichiometrically.

Interest in the degradation was initialised when, upon opening a dispersion of $TiS_2$, a sulphurous smell was noted. In the presence of water $TiS_2$ will oxidise according to the following reactions:

$TiS_2 + xH_2O \rightarrow TiS_{2-x}O_x + xH_2S_2$

whilst in oxygen:

$TiS_2 + xO_2 \rightarrow TiS_{2-x}O_x + xS_2$ is likely.

Dolui *et al*. performed a series of DFT calculations to predict that in an aqueous environment, water would preferentially oxidise a monolayer from the edge or from a point defect if the edge was 50% terminated in sulphur. This correlated well with work by Han *et al*[14] who observed similar behaviour in $TiS_2$ nanocrystals in a water/toluene mixture. We aimed to expand this work to study the oxidisation of material produced through liquid phase exfoliation methods on two fronts: *Ex-situ* aging experiments to study the behaviour of $TiS_2$ stored under deionised (DI) water, and *in-situ* heating experiments to study the behaviour of $TiS_2$ in an oxygen/argon mixture with varying temperature.

Results and Discussions

Flakes exfoliated by ultra-sonication had thicknesses in the approximate range 10-60 nm as measured by EFTEM thickness mapping, taking the electron inelastic mean free path at 300 kV to be 127.4 nm[15] (further information can be found in the supplementary information). Figure 1 shows that flakes appeared to be of fairly uniform contrast in phase contrast TEM and crystalline apart from a small amorphous region around the edge of each flake, seen in Figure 1B. Mapping the oxygen K-edge with EELS (Figure 1F) shows this is due to oxidisation of the edge of an otherwise fairly pristine $TiS_2$ flake, similar to $TiS_2$ flakes studied with High Resolution STEM (Dolui *et al*). The IPA dispersion clearly had some effect on the flakes; those drop-cast one month later showed that the edge oxidisation had grown approximately 20nm into the flake (Figure 2 A shows the change in STEM contrast at the edge of one flake, whilst Figures 2B-F demonstrate the reduction in sulphur and the increase in oxygen at this edge region, leaving the titanium relatively unchanged.). This is most likely the degradation occurring in the original dispersion.

Figures 3 B-F describe how, over the course of a further 21 days in DI water, this oxidisation layer continued to grow into the flake, with notable variations in contrast in the HAADF STEM image from edge to centre of the sheet (Figure 3 A). The EDX maps of sulphur and oxygen (Figures 3 B and E) closely match the boundary between these regions, demonstrating oxidisation in agreement with the mechanism predicted. It should be noted that oxygen appears to be present over the entire width of the flake. It is not clear if this is due to some degree of surface oxidisation or due to the absorption to the surface of water molecules, however we can clearly see the change in morphology of the flake by looking at the diffractograms in Figure 3 (1-4). In order to obtain highly localised crystallography information, a microprobe ~15 nm diameter was used to obtain diffraction patterns from across the flake. The sampled area is therefore over 130 times smaller than that obtained with our smallest selected area electron diffraction (SAED) aperture. Figure 3 G-J show that the oxide region is amorphous whereas the sulphurous region is still crystalline. Other, smaller, flakes were observed to be fully oxidised within this time period.

The way the flakes oxidise in water (inwards, from the edge), matches the way flakes have been observed to degrade within a solvent dispersion. Since water is miscible in IPA, it would be reasonable to conclude that small amounts of water contaminating the solvent were responsible for the slow degradation observed.

Another potential route to oxidisation would be from oxygen gas. We studied the response of $TiS_2$ flakes to exposure to a 20% oxygen/argon mixture between 150 and 500 °C using the same combination of EDX and EELS spectrum imaging and microprobe line profiles, in conjunction with a heating/gas in-situ TEM holder. To try and reduce any initial oxidisation, the flakes were dispersed in anhydrous IPA in the inert atmosphere of a nitrogen glove-box. The sample was kept under vacuum within a few hours of being drop-cast. The initial state of the flake is shown in Figure 4 A-F. Again, we observed a small degree of oxidisation around the edges of the flake, particularly apparent from the O-K edge EELS map in Figure 4C. However, we also observed contamination starting to build up around the edges of flakes, containing both silicon and oxygen. As expected, we did not observe any changes to the flakes immediately upon introducing the gas mixture with the holder temperature at 150°C. The holder temperature was subsequently increased to 250°C with no apparent change in flake integrity. The next heating increment was to 375°C, immediately inducing a change in one flake. The temperature was subsequently dropped to 325°C to prevent further changes whilst this one flake was investigated.

It was apparent that the morphology of the flake had completely changed – a previously uniform flake now appeared to have striations running through it. Figure 5 compares this flake before and after, and it can clearly be seen that the main flake has lost almost all the sulphur. However the flake directly above it still appears to be intact. This behaviour was observed across the entire sample, some flakes would appear to have oxidised almost immediately, whilst others remained unchanged. Figure 6 shows a flake, to the top-left of the main one in Figure 4 A, which had partially oxidised at the centre causing sulphur to be removed from the centre. This hole was seen to grow during the acquisition of the EDX spectrum image (Figure 6B), most likely due to electron-beam induced sputtering, and then further when acquiring EELS data (Figure 6F). We observed the oxide to be present across the entirety of the acquisition region, even at the top right which appeared unchanged between Figures 6B and F. This possibly suggests that the flake had mostly oxidised whilst there was still some sulphur detected on the surface, which then was subsequently removed by the beam rather than an ongoing oxidisation reaction. HRSTEM imaging of oxidised regions across the specimen showed clear polycrystalline domains, in agreement with some microprobe diffraction data taken from some of the oxidised flakes.

Analysis of structure from Electron Energy-Loss Spectra

Electron energy-loss spectra contain information beyond merely the elemental composition of a sample. Both the position and shape of an edge can contain information pertinent to a range of properties – depending on the specific transition occurring. Core loss edges are produced upon excitation of core-shell electrons, and can tell one much about environment of the source atom.

For the sample oxidised *ex*-situ, there appears to be a shift of ~3.5 eV on the spectrum with regards to the expected onset of the Ti $L_{2,3}$ edge. The oxygen K edge has two features in both the 'oxide' and 'sulphide' regions. Even accounting for the shift, the first appears to be higher energy than the pre-peak observed in the in-situ sample. However, since the same features were observed across both regions, it would be reasonable to conclude there was some degree of surface oxidisation across the specimen. The amorphous nature of the oxide could explain the lack of splitting of the Ti $L_{2,3}$ edges, which would be expected in the crystalline rutile and anatase structures (Supplemental figure 3). However, it is extremely challenging to identify the exact stoichiometry of an amorphous region through comparison with white line Energy Loss Near Edge Structures (ELNES). Figure 7A – integrated Ti spectra from Figure 3 over oxide region.

From the *in-situ* sample, however, we were able to distinguish between oxygen bound in titanium oxide and that present in either molecular form or within the contamination region, identifying oxygen bound to titanium through an O-K edge pre-peak, using energy windows of 528.5-532.1 eV and 528.1 – 538.7. This is attributed to the interaction between hybridised O 2p orbitals with the Ti 3d orbitals[16]. Similarly to the ex-situ oxide regions, no splitting was observed on the Ti $L_{2,3}$ edges, despite evidence that the oxide is crystalline. Excluding rutile and anatase from the possible structures would imply that the oxide formed is not $TiO_2$, but something else. The shoulder on both the $L_3$ and $L_2$ edges matches those observed on lower valence oxides such as $Ti_5O_9$ and $Ti_4O_7$. Fig 7B – integrated Ti spectra from part of flake in Figure 6.

Looking at the relative intensities of the $L_2$ and $L_3$ peaks gives some idea of the valence of the titanium: The greater $I(L_2)/I(L_3)$, the closer the valence is to 4+. According to the empirical fit put forward by Stoyanov et al, our ratio of 1.040 would suggest a $Ti^{4+}$ concentration (defined as $Ti^{4+}/\sum Ti$) of -0.0686 - below the realistic detection limits of 1 at%. Despite difference is shape therefore, we should conclude the oxide formed to realistically be $Ti_2O_3$, where the valence is 3+. Similar calculations for the oxide region in the hydrated sample give $I(L_2)/I(L_3)$ = 1.102, for a concentration of $2.51 \times 10^{-3}$. Again this is too low to be seen as accurate, but suggests the valence to be around 3+.

Density Functional Theory

We have also shown that it might be possible to have some degree of control over the electronic properties of a sulphide-oxide hybrid material. Oxidisation by water is a sufficiently slow process that it might be possible to generate flakes of a desired composition by controlled exposure to water for a desired length of time. Calculations have shown the band-gap can be fine-tuned between the extremes of pure $TiS_2$ and $TiO_2$ via oxygen substitution[17], which might also explain why such a range of values exist for the measured band-gap.

More recent calculations have been performed with regards to the thermodynamics of $O_{2\,(g)}$ oxidisation. They have shown that the filling of a sulphur vacancy by oxygen is strongly exothermic (+3.39 eV), whilst substitution at of sulphur for oxygen is exothermic both the edge (+2.00 eV) and centre (+1.98 eV) of a flake. The physisorption of an oxygen gas molecule onto the edge is endothermic (-0.28 eV) whilst for a water molecule it is exothermic (+0.76 eV). We would therefore expect the limiting factor for the reaction with oxygen gas to be a kinetic barrier

Experimental Methods

Commercial $TiS_2$ powder was acquired from Sigma Aldrich (99.9% Lot#333492) and stored in an argon filled glove box with both oxygen and moisture readings < 100 ppm. 5 mg of $TiS_2$ was added to 50 ml anhydrous IPA (99.5 % Sigma Aldrich Lot#278475). The unexfoliated dispersion was then exfoliated using a Fisherbrand ultrasonic disintegrator operating at 20 kHz and 34W. The temperature was kept at a constant 10°C using a circular IsoUK 4100 R20 refrigeration bath. The dispersion was sonicated for three hours and quickly placed inside a centrifuge tube. The dispersion was centrifuged at 1000 RPM for 1 hour and the supernatant was decanted and stored in the Argon filled glove box.

For the *ex-situ* ageing experiments, the IPA dispersion was drop-cast onto gold TEM grids, whilst for the *in-situ* experiments anhydrous IPA dispersion was drop-cast onto specialised Si-$Si_3N_4$ chips designed to enable both local heating and to prevent a gas mixture from reaching the vacuum inside the microscope.

The microscope used for the *ex-situ* work was an FEI Titan 80-300 S/TEM, operated mainly in scanning transmission electron microscopy (STEM) mode at 300kV, with an EDAX energy dispersed X-ray (EDX) detector and a Gatan Imaging Filter (GIF) for recording electron-energy loss spectra (EELS).

The *in-situ* work was performed on an FEI Titan G2 80-200 S/TEM (ChemiSTEM), with SuperX EDX detectors and a Quantum GIF. We used a Protochips Atmosphere gas environmental cell holder with an oxygen/argon mix at 290-350 mbar operated between 150-500 °C. The holder had been modified previously, similar to the modifications reported for the liquid-cell holder[18], in order to significantly reduce the shadowing of the holder on two of the four available EDX detectors.

The procedure for *ex-situ* hydration of $TiS_2$ was to initially image the flakes as soon as possible after dispersion, then to store the grid in DI water in between imaging sessions. The grid was dried in an Across International drying oven (V0-16020C3-Prong) at ~$5\times10^{-2}$ mbar for an hour prior to imaging to remove as much excess water from the grid as possible, and then returned to DI water immediately afterwards. Attempts were made to re-image the same flakes, but build-up of carbonaceous contamination would coat imaged areas, retarding or fully preventing further oxidisation. However, the flakes studied were broadly similar in both size and thickness, so reasonable comparisons could be drawn between them.

Density functional theory[19,20] calculations were performed with the VASP[21] and CP2K[22] packages using the PBE (Perdew-Becke_Ernzerhof) exchange and correlation (XC) functional. A PAW (projected augmented-waves) approach was used with a cut-off energy of 400 eV, sampled with a 5x5x1 Monkhorst-Pack *k*-grid.

Conclusions

The main result to be interpreted from these results is the kinetic barrier for oxygen-gas-based oxidisation appears to be significantly higher than that for oxidisation via water. Water was observed to slowly oxidise $TiS_2$ flakes, initiating at the flake edge, forming an amorphous oxide, with EELS suggesting a Ti valence of around 3+. Oxygen gas was only observed to oxidise within an in-situ set-up, above 375°C, whereupon flakes oxidised immediately to form poly-crystalline material, also with a Ti valence of around 3+. DFT calculations have shown the oxidisation of $TiS_2$ by both water and oxygen gas to be thermodynamically favourable.

Acknowledgements

EI, SO'B and VN wish to thank the support of the SFI PIYRA grant, the European Research Council (2DNanoCaps project) and the EU ITN MoWSeS project within the FP7 framework. The Advanced Microscopy Laboratory (AML) and its staff (in particular CD) are thanked for their assistance in electron microscopy in Dublin. SS and CSC have been supported by the European Research Council (Quest-project). All calculations were performed on the Parsons cluster maintained by the Trinity

Centre for High Performance Computing, under project id: HPC_12_0722. This cluster was funded through grants from Science Foundation Ireland.

EL wrote the first draft of the paper, and performed the ex-situ experiments. EAL, EP, SJH, CD and EL performed the in-situ experiments. CSC and SS performed simulations of the oxidisation. SO'B prepared the dispersions. VN and EL planned the experiments.

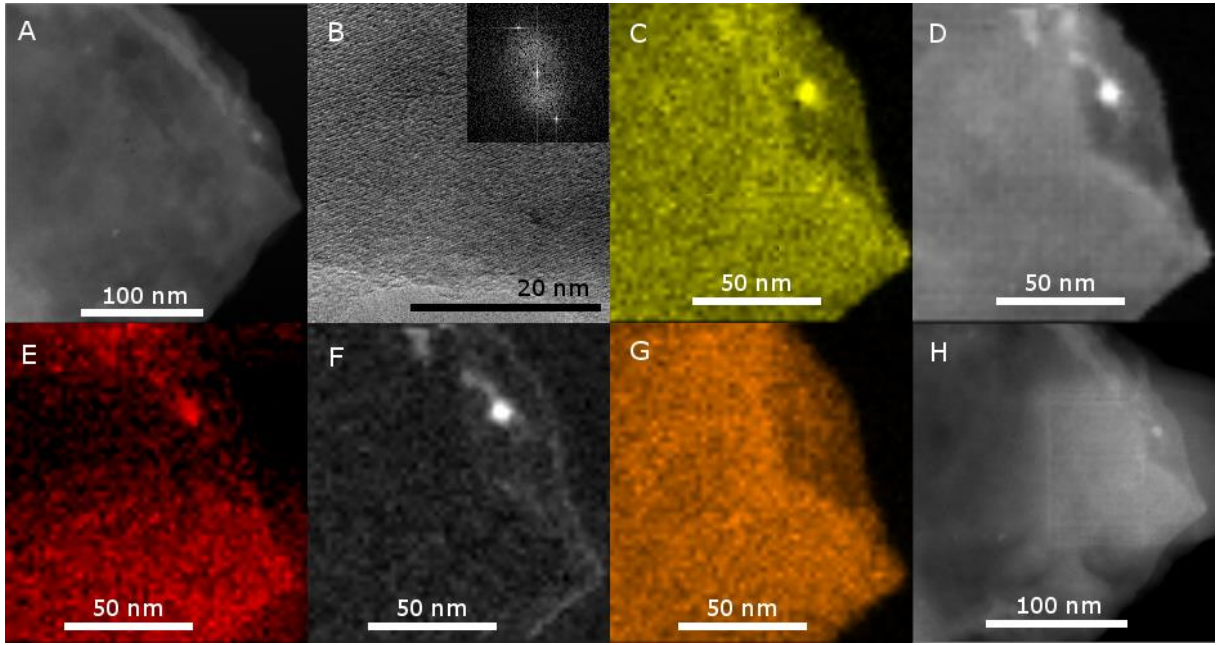

**Figure 1:** Initial status of TiS2. A: STEM image of SI region, B: TEM of edge of flake with FFT inset, C: Ti-K EDX map, D: Ti-L2 EELS map, E: O-K EDX map, F: O-K EELS map, G: S-K EDX map, H: STEM of region after SI acquisition

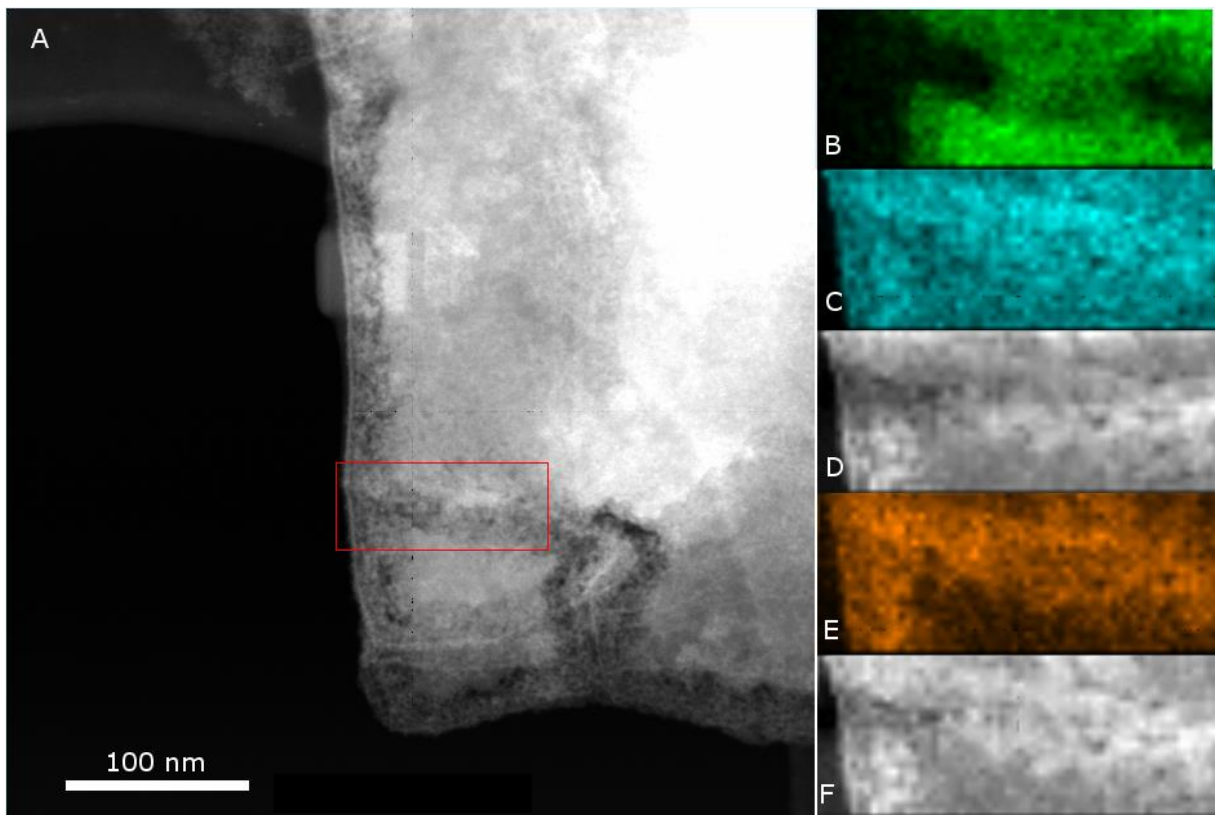

**Figure 2:** Slightly degraded TiS2. A: STEM image of degradation at edge of TiS2 flake, B: S-K EDX map, C: Ti-K EDX map, D: Ti-K EELS map, E: O-K EDX map, F: O-K EELS map

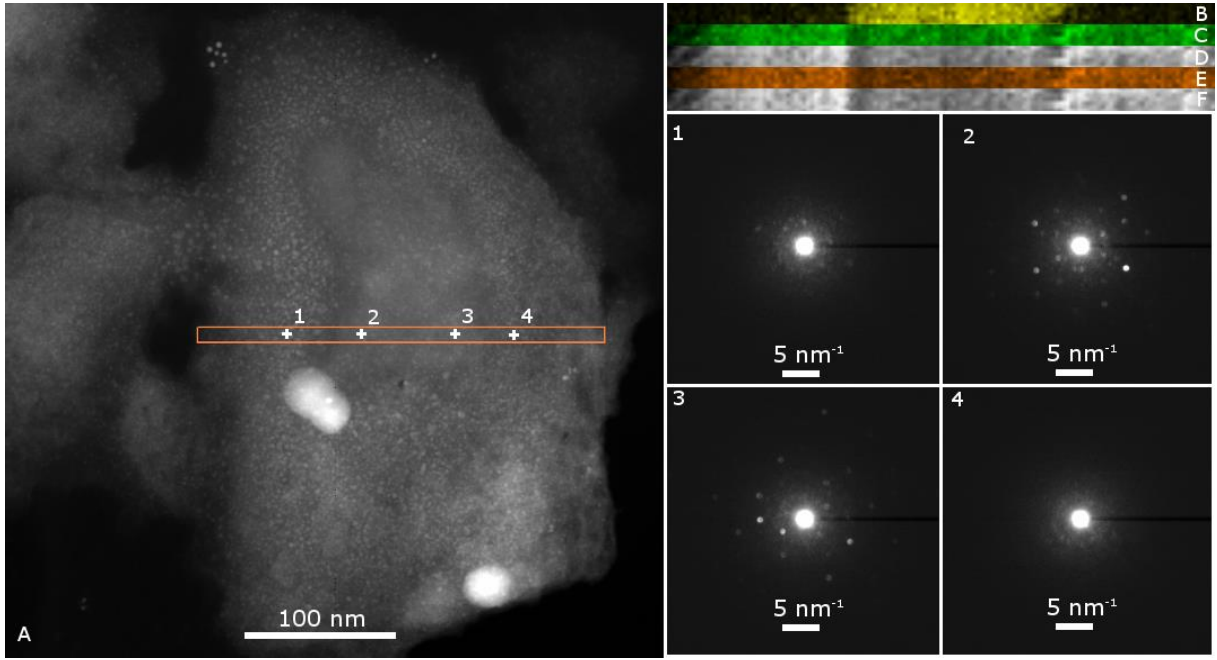

Figure 3: Degradation of TiS2 flake after 22 days in DI water A: DF STEM Overview of flake, B: S K-edge EDX Map, C: Ti K-edge EDX Map, D: Ti L2 EELS Map, E: O K-edge EDX Map, F: O K-Edge EELS Map, 1-4 microprobe diffraction pattern from corresponding position

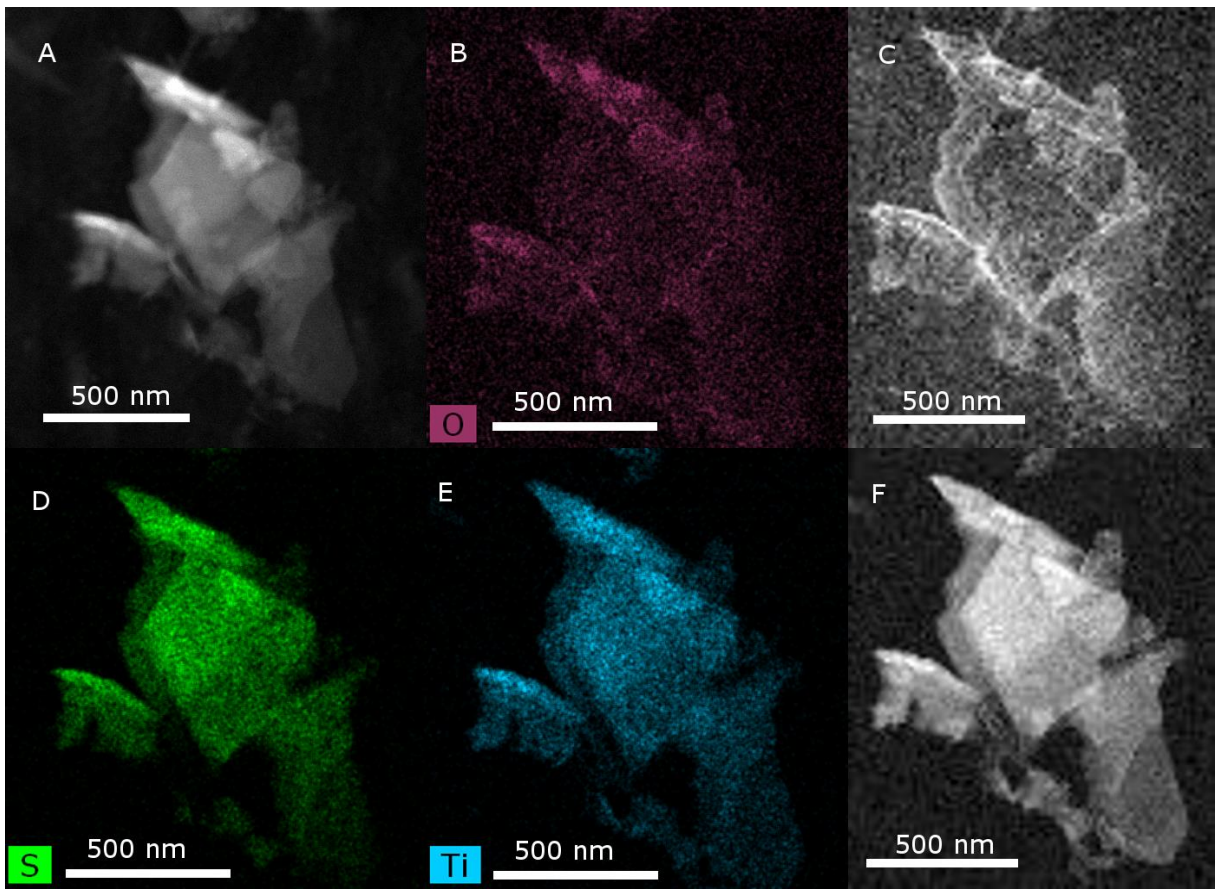

Figure 4: Initial status of in-situ TiS2. A: HAADF Image, B: O K EDX map, C: O K EELS map, D: S K EDX map, E: Ti K EDX map, F: Ti L2 EELS map

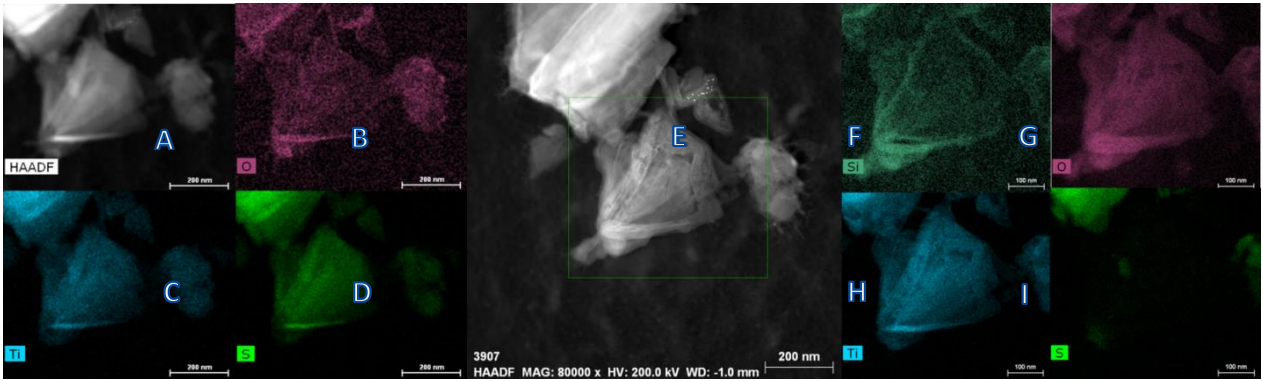

**Figure 5: Fully oxidised TiS2 flake. A: Initial HAADF, B: O K EDX map, C: Ti K EDX map, D: S K EDX map, E: Post-oxidisation HAADF, F: Si K EDX map, G: O K EDX map, H: Ti K EDX map, I: S K EDX map.**

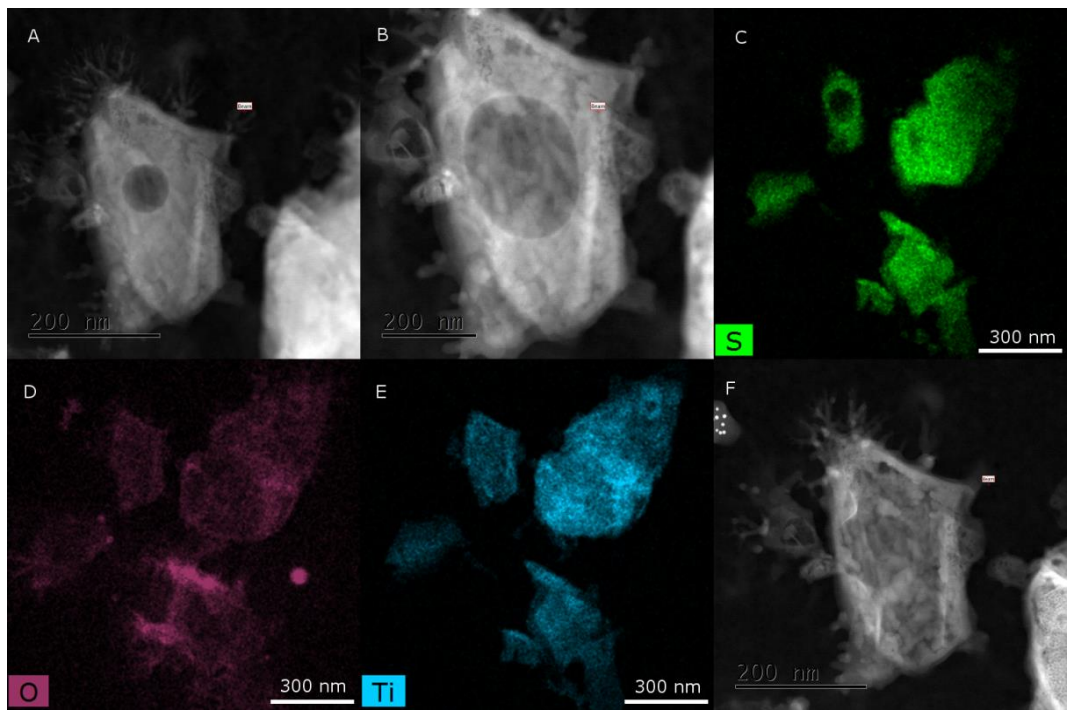

**Figure 6: Loss of sulphur during specta acquisition. A: HAADF of partially oxidised flake pre-EDX acquisition, B: HAADF of flake post-EDX acquisition, C: S K EDX map, D: O K EDX map, E: Ti K EDX map, F: HAADF post-EELS acquisition**

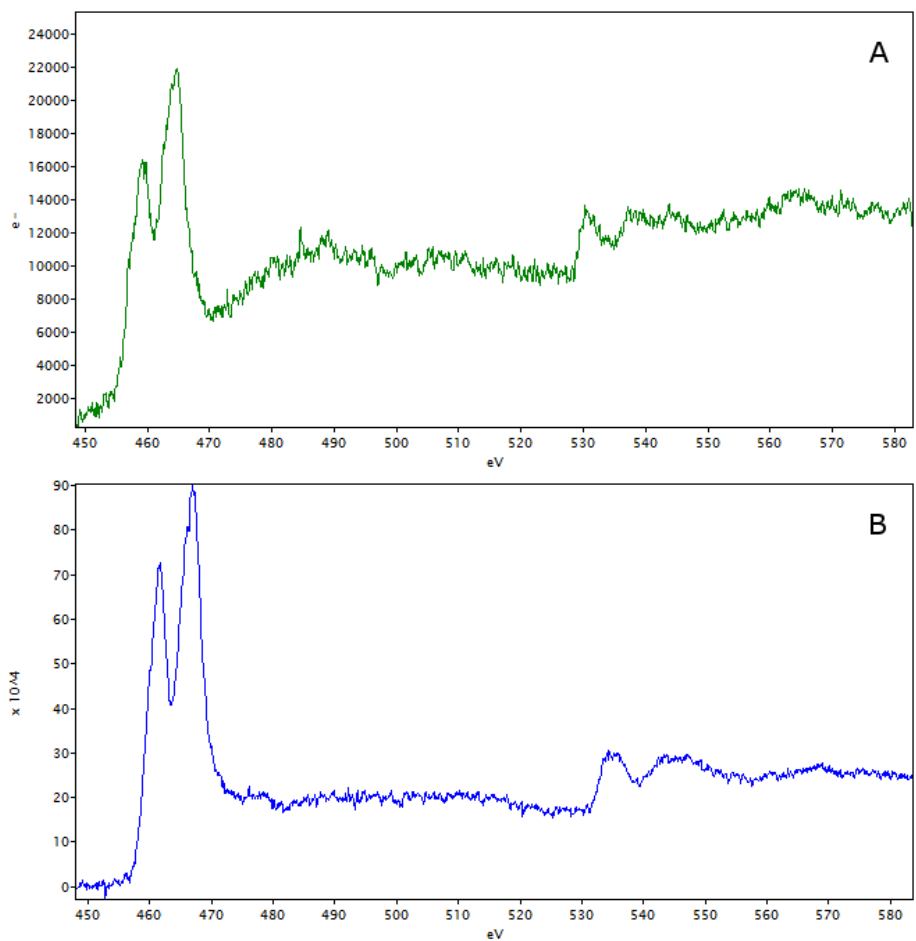

Figure 7: Ti L$_{2,3}$ and L$_1$ edges from A: *ex-situ* hydrated TiS$_2$, B: *in-situ* oxidised TiS$_2$

Supplementary Information

In STEM, electrons are accelerated to 300 kV before being focussed into a sub-nanometre probe. The probe is rastered across a region of interest and a range of signals collected at each pixel. Annular bright field (ABF) and high angle annular dark field (HAADF) images are formed by plotting the intensity of the signal scattered to each annulus at each point. The ABF detector collects electrons scattered through a small angle, whilst the HAADF collects those scattered to 15-30 rad (with the intensity proportional to $Z^{1.67}$). Each interaction between beam and sample will result in the primary beam losing energy corresponding to that interaction. Some of these interactions are element specific (for example the excitation of core-shell electrons), and these can indicate the chemical state of that atom (EELS). These excitations can also generate X-rays specific to the originating atom, which are collected in EDX.

Thickness mapping in EFTEM is performed according to the EELS Log-Ratio method put forward by Malis et al[15], giving thickness to ±20% in inorganic specimens. The local thickness, t, is given by:

$$t = \lambda \ln\left(\frac{I_t}{I_0}\right)$$

Where $I_t$ and $I_0$ are the integrated intensities of the total and zero loss peak in an EEL spectrum, and λ is the inelastic mean free path of a primary beam electron. λ can be determined as being parameterised on three factors: the atomic number Z of the scattering material, the incident beam energy, $E_0$ in keV, and β, the spectrum collection semiangle in mrad.

$$\lambda = 106\, F \frac{(\frac{E_0}{E_m})}{\ln(2\beta \frac{E_0}{E_m})}$$

Where $= \frac{1+\frac{E_0}{1022}}{\left(1+\frac{E_0}{511}\right)^2}$, , and $E_m = 7.6\, Z^{0.36}$.

The intensities $I_0$ and $I_t$ are designated as the following parts of a standard EEL spectrum (Supplementary Figure1) :

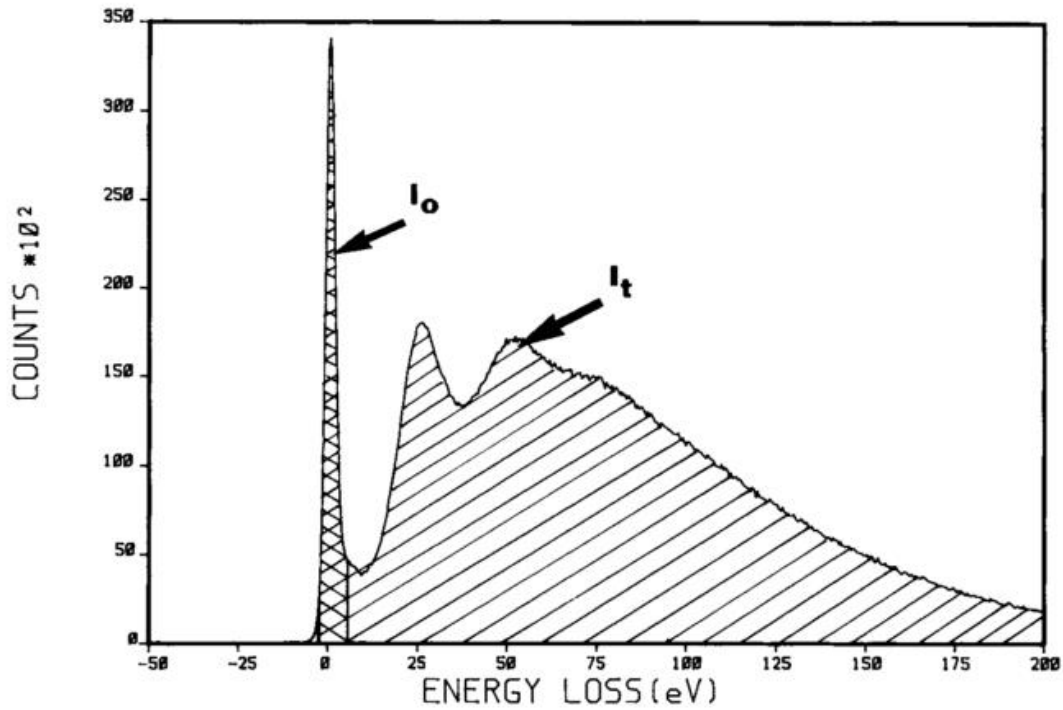

Fig. 1. Typical energy-loss spectrum from a 400-nm carbon film showing $I_t$ and $I_0$ in equation (1).

Supplementary Figure 1: Designation of $I_0$ and $I_t$ in a typical electron energy-loss spectrum[15].

For TiS$_2$, with an effective Z of 18.13, λ for a 300keV electron is determined to be 127.4 nm. An example of the formation of a thickness map is shown in Supplementary Figure 2:

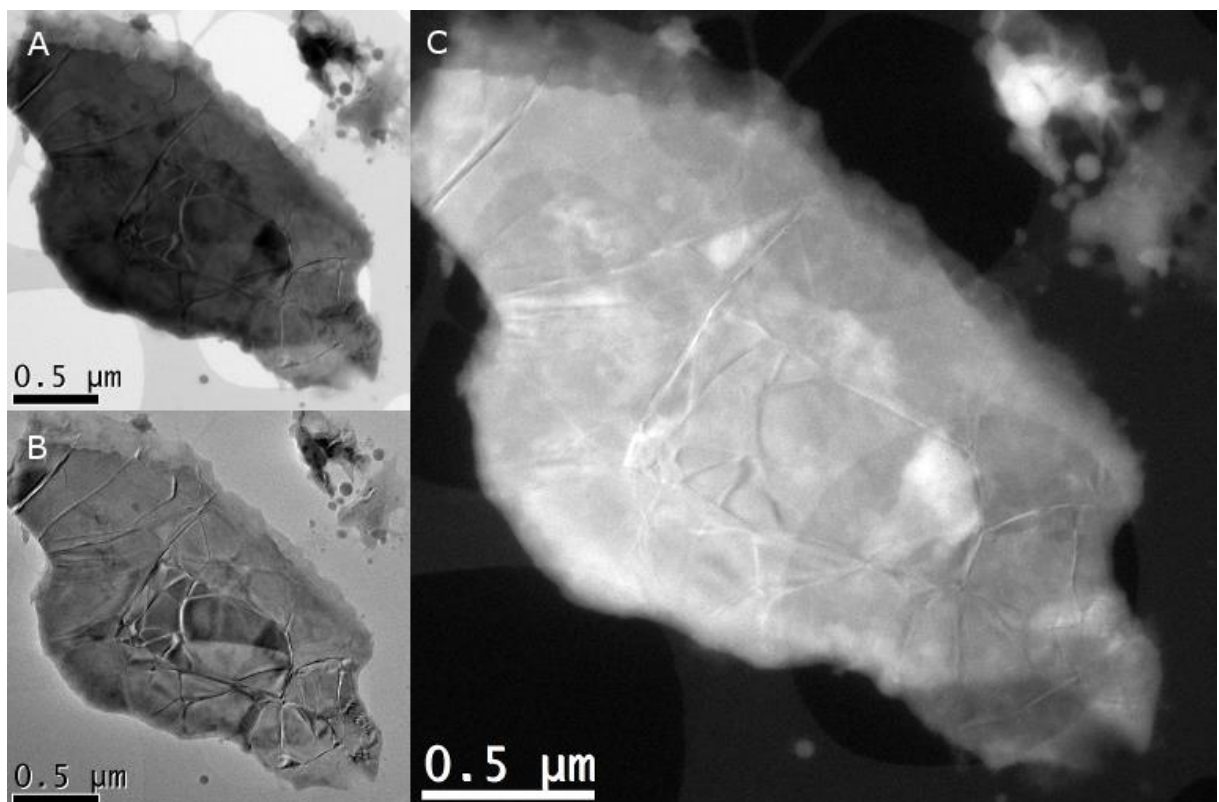

Supplementary Figure 2 A: Elastic ($I_0$) EFTEM image, B: Inelastic ($I_t$) EFTEM image, C: Processed thickness map

whilst the inelastic image (supplementary figure 2B) is formed from the entire energy range (204.8 eV). The thickness map is then formed by multiplying the image formed by $\ln\frac{I_t}{I_0}$ by λ (127.4 nm), such that the intensity of each pixel is the thickness at that point in the sample (for the regions of flake suspended over vacuum).

Reference spectra for various stoichiometries of titanium oxide were taken from the 2007 work of Stoyanov *et al.*[16], reproduced below in supplementary figure 3:

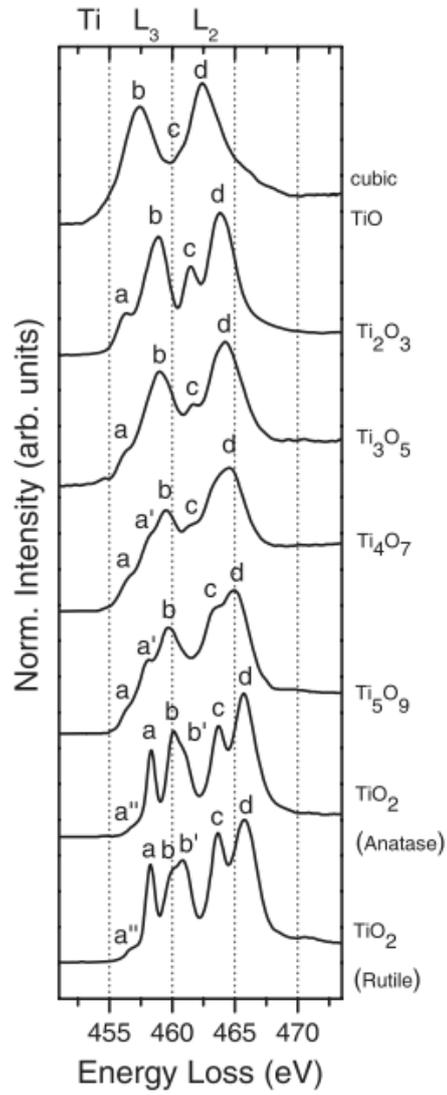

**FIGURE 5.** Ti $L_{3,2}$ ELNES spectra of the analyzed Ti oxides. The valence state of Ti decreases from the bottom to the top.